\documentclass[usenatbib,usegraphicx,letterpaper]{mn2e}
\usepackage[totalwidth=480pt,totalheight=680pt]{geometry}

\usepackage{amssymb}
\usepackage{epsfig}
\usepackage{amsmath}
\usepackage{color}

\newcommand{\apjl}{ApJL~}
\newcommand{\apjs}{ApJS~}

\newcommand{\unit}[1]{\mathrm{#1}}

\newcommand{\wpp}{w_{\mathrm{p}}}
\newcommand{\wprp}{w_{\mathrm{p}}(r_{\mathrm{p}})}
\newcommand{\rp}{(r_{\mathrm{p}})}

\newcommand{\kms}{\unit{km \ s^{-1}}}

\newcommand{\hmpc}{h^{-1}\mathrm{Mpc}}
\newcommand{\hmpcvol}{h^{-3}\mathrm{Mpc^{3}}}
\newcommand{\hkpc}{h^{-1}\mathrm{kpc}}
\newcommand{\hMsun}{h^{-1}M_{\odot}}

\newcommand{\Mvir}{M_\mathrm{vir}}

\newcommand{\Rvir}{R_\mathrm{vir}}

\newcommand{\Msun}{M_{\odot}}

\newcommand{\pimax}{\pi_\mathrm{max}}

\newcommand{\vmax}{\mathrm{V}_\mathrm{max}}

\newcommand{\vpeak}{\mathrm{V}_\mathrm{peak}}

\newcommand{\mchar}{M_{\mathrm{char}}}
\newcommand{\mhost}{M_{\mathrm{host}}}

\newcommand{\zquench}{z_{\mathrm{quench}}}
\newcommand{\zform}{z_{\mathrm{form}}}
\newcommand{\zacc}{z_{\mathrm{acc}}}
\newcommand{\zchar}{z_{\mathrm{char}}}
\newcommand{\zstarve}{z_{\mathrm{starve}}}

\newcommand{\tdelay}{t_{\mathrm{delay}}}

\newcommand{\veff}{V_{\mathrm{eff}}}

\newcommand{\Psdss}{P_{\mathrm{SDSS}} ( g-r | L )}

\newcommand{\fblue}{F_{\mathrm{BLUE}} ( L )}

\newcommand{\beq}{\begin{equation}}
\newcommand{\eeq}{\end{equation}}
\newcommand{\beqray}{\begin{eqnarray}}
\newcommand{\eeqray}{\end{eqnarray}}

\newcommand{\ben}{\begin{enumerate}}
\newcommand{\een}{\end{enumerate}}
\newcommand{\bit}{\begin{itemize}}
\newcommand{\eit}{\end{itemize}}

\usepackage{epsfig}  \usepackage{graphicx}   \usepackage{rotating}

\begin{document}

\title[The Dark Side of Galaxy Color]
{The Dark Side of Galaxy Color}

\author[A. P Hearin \& D. F Watson]
{Andrew P. Hearin$^1$ and
  Douglas~F.~Watson$^2,^3$ \\
$^1$Fermilab Center for Particle Astrophysics, Fermi National Accelerator Laboratory, Batavia, IL, 60510-0500 \\
$^2$NSF Astronomy \& Astrophysics Postdoctoral Fellow,
  Department of Astronomy \& Astrophysics, The University of Chicago,
  Chicago, IL 60637 \\
$^3$Kavli Institute for Cosmological Physics, 5640 South
  Ellis Avenue, The University of Chicago, Chicago, IL 60637}

\maketitle

\begin{abstract}

We present {\em age distribution matching}, a theoretical formalism for predicting how galaxies of luminosity L and color C occupy dark matter halos. Our model supposes that there are just two fundamental properties of a halo that determine the color and brightness of the galaxy it hosts: the maximum circular velocity $\vmax,$ and the redshift $\zstarve$ that correlates with the epoch at which the star formation in the galaxy ceases. The halo property $\zstarve$ is intended to encompass physical characteristics of halo mass assembly that may deprive the galaxy of its cold gas supply and, ultimately, quench its star formation. The new, defining feature of the model is that, at fixed luminosity, galaxy color is in monotonic correspondence with $\zstarve,$ with the larger values of $\zstarve$ being assigned redder colors. We populate an N- body simulation with a mock galaxy catalog based on age distribution matching, and show that the resulting mock galaxy distribution accurately describes a variety of galaxy statistics. Our model suggests that halo and galaxy assembly are indeed correlated. We make publicly available our low-redshift, SDSS $M_{r}<-19$ mock galaxy catalog, and main progenitor histories of all $z=0$ halos, at http://logrus.uchicago.edu/$\sim$aphearin

\end{abstract}




\section{INTRODUCTION}
\label{sec:intro}


Galaxies are not color blind in how they congregate with one another.
Redder galaxies (older with little active star formation) tend to occupy more
dense environments, and conversely, bluer galaxies (younger with
active ongoing star formation) preferentially reside in underdense
regions
\citep{balogh99,blanton05,weinmann06b,weinmann09,peng_etal10,peng_etal12,wetzel_etal11,carollo_etal12}.
The color dependence of galaxy location within the cosmic web also manifests in measurements of two-point statistics 
\citep[e.g.,][]{norberg02,zehavi02,zehavi05a,li06,yang11b,zehavi11,mostek12}.  
Additionally, there is a clear bimodality in the
distribution of galaxy color and brightness, with distinct red and blue sequences in
color-magnitude space, along with a minority 'green valley' population
\citep{blanton03,baldry04,blanton05,wyder07}. This segregation between blue
star-forming and red passively evolving galaxies has been shown to
persist out to $z\sim1$ \citep{bell04,cooper06,cooper12} and even as
high as $z\sim3$ \citep{whitaker11}.  While these correlated trends in color-magnitude and configuration space are observationally well-established,
our theoretical understanding of the physical processes driving these correlations remains incomplete.

In the standard $\Lambda$CDM cosmological model structures assemble
hierarchically, wherein high density regions condense and virialize,
forming bound structures known as dark matter halos.  Halos of
sufficient mass are the natural sites for luminous galaxies to form
\citep{whiterees78,blumenthal_etal84}, and the way galaxies connect to
dark matter halos provides the fundamental link between predictions of
galaxy formation theory, such as color, and the concordance
$\Lambda$CDM model. Understanding this link requires a detailed knowledge of the spatial
distribution of galaxies and halos, and towards this end large galaxy redshift surveys
such as the Sloan Digital Sky Survey \citep[SDSS:][]{york00a}  have
produced detailed three-dimensional maps of galaxies in the local
Universe. 
 
Extensive effort has been put forth to understand how, as a function
of color, galaxies map to dark matter halos in a cosmological context,
and there are two general approaches to constructing the galaxy-halo
connection. First, there is the halo occupation distribution (HOD)
technique (and the closely related conditional luminosity function
technique \citep[CLF: e.g.,][]{yang03,vdBosch07}), which adopts
parametric forms for the biased relation between
galaxies and halo mass \citep[e.g.,][]{peacock00a, seljak00,
  scoccimarro01a, berlind02, cooray02, zheng07, leauthaud11a,watson_etal12a}.   The
HOD describes this relation by specifying the probability distribution  
that a halo of mass $M$ contains $N$ galaxies, along with a
prescription for the spatial distribution of galaxies within halos.
In this formalism, it is typically assumed that a single
'central' galaxy lives at the center of each distinct halo, with additional
'satellite' galaxies tracing the dominant dark matter component. The
most common implementations of this approach do  not explicitly use
any information about dark matter subhalos (self-bound, dark matter
structures orbiting within their host halo potential). 

The second technique is known as abundance matching
\citep{kravtsov04a, vale_ostriker04, tasitsiomi_etal04,
  vale_ostriker06, conroy06, conroy_wechsler09, guo10, simha10,
  neistein11a, watson_etal12b, reddick12, rod_puebla12,
  hearin_etal12b,kravtsov13}.  This is a simple, yet remarkably
powerful approach wherein by assuming a monotonic relation between
maximum circular velocity $\vmax$ (or halo mass) and luminosity $L,$
(or stellar mass) one can `abundance match' to make the correspondence
between a dark matter (sub)halo and the galaxy residing at its center.
 In this case, galaxies are assigned to dark matter halos
with the most massive galaxies residing in the most massive halos.
This yields an implicit relationship between $L$ and $\vmax$ such that, by construction, the luminosity function of 
mock galaxy catalogs is in exact agreement with the data.

The most accurate and precise theoretical descriptions of the galaxy color-halo connection to date have been
predominantly derived from modeling the color-dependent galaxy
two-point correlation function (2PCF) through the HOD formalism
\citep[e.g.,][]{zehavi05a,zehavi11,skibba_sheth09,tinker_wetzel10,krause_etal13}, or semi-analytic models \citep{white_frenk91,kauffmann_etal99}. 
However, abundance matching type theories implemented in a cosmological N-body simulation are now being explored \citep{gerke12,masaki13}.  Alternatively, new phenomenological-based approaches that seek to simplify the complexities of semi-analytical models are actively being pursued \citep{peng_etal10,mutch_etal13,lilly_etal13}.  

Our intent in this paper is to try and uncover the simplest,
physically-motivated model for  associating the color of a galaxy with
the mass accretion history of the galaxy's parent halo. To that end,
we introduce a new theoretical formalism which we call {\em age distribution matching}.  This model supposes that there are just
two fundamental properties of a halo that determine the color and
brightness of the galaxy it hosts, (1) the maximum circular velocity of a test particle orbiting in its
gravitational potential well (i.e., $\vmax,$), and (2) the redshift at
which the galaxy will be starved of cold gas, which we refer
to as $\zstarve$ throughout the paper.  This halo property $\zstarve$
is intended to encompass characteristics of a halo's assembly history
which should be associated with physical processes that can deprive a
galaxy of a cold gas supply and, ultimately, quench star formation. The novel feature of the model is that,
at fixed luminosity, galaxy color is in monotonic correspondence with
$\zstarve.$  By constructing a mock galaxy catalog with
both $r$-band magnitudes (luminosities) $M_r$ and $(g-r)$ colors, we compare the predictions of our model to a number of
observed galaxy statistics.

The rest of this paper is organized as follows.  In \S~\ref{sec:data}
we discuss the data incorporated throughout this work and in
\S~\ref{sec:sims} we discuss the simulation, halo catalogs, and merger trees used in
this study.  In \S~\ref{sec:mocks} we describe our algorithm for
constructing our SDSS-based mock catalog including a detailed
description of our age distribution matching formalism.  Results
are presented in \S~\ref{sec:results} followed by a discussion in
\S~\ref{sec:discussion}.  Finally, in \S~\ref{sec:conclusion} we give
a summary of our work.  Throughout we assume a flat $\Lambda$CDM
cosmological model with $\Omega_{\mathrm{m}}=0.27$ and all quoted
magnitudes use $h=1$.


\section{Data}
\label{sec:data}

We construct our low-redshift mock galaxy sample based on the galaxy
catalog described in \S~\ref{subsec:galaxysample}. In particular, this
is the sample that defines the distribution of colors at fixed
luminosity, $\Psdss,$ required in age distribution matching (see
\S~\ref{subsubsec:grassign}). In \S~\ref{subsec:clustering} we describe the
projected 2PCF measurements that we
use to compare with the clustering predictions of our mock sample. We
use the galaxy group sample described in \S~\ref{subsec:sdssgroups} to
test our predictions for how galaxies arrange themselves into groups
by color. 

\subsection{SDSS Galaxy Sample}
\label{subsec:galaxysample}

For our baseline galaxy sample we use a volume-limited catalog of
galaxies constructed from the Main Galaxy Sample of Data Release 7
(\citet{DR7_09}, DR7 hereafter) of SDSS. This galaxy catalog is an
update of the \citet{berlind06} sample, which was based on SDSS Data
Release 3, to which we refer the reader for details. Briefly, our
volume-limited spectroscopic subsample
($\veff=5.8\times10^{6}\hmpcvol$) spans the redshift range $0.02 \leq
z \leq 0.068$ with $r$-band absolute magnitude $M_{r} - 5\log h <
-19.$ For $M_{r},$ we use Petrosian magnitude measurements. Throughout this paper we refer to this as the ``Mr19" galaxy
sample.

The term ``fiber
collisions" refers to cases when the angular separation between two or
more galaxies is closer than the minimum separation ($55''$) permitted by the
plugging mechanism of the optical fibers in the spectrometer (see
\citet{guo_zehavi_zheng12}, and references therein). The treatment of
fiber collisions in the DR7-based galaxy sample we use in this paper
differs from that in the catalog presented in \citet{berlind06}. These
differences were discussed in detail in Appendix B of
\citet{hearin_etal12b}, to which we refer the reader for details. 

\subsection{Clustering Measurements}
\label{subsec:clustering}

We compare our model predictions for the color-dependent 2PCF to the
SDSS measurements presented in \citet{zehavi11}. Using the observed
$(g-r)$ colors,  they separate red and blue galaxy populations based
on the following magnitude-dependent color cut: \beq
\label{eq:colorcut}
(g-r) = 0.21- 0.03M_r.   
\eeq  
We use the luminosity-binned clustering
data presented in their Tables C9 \& C10 to test our model 2PCF
predictions for red and blue galaxies. 

\subsection{Galaxy Group Measurements}
\label{subsec:sdssgroups}

Galaxy groups are constructed from the Mr19 galaxy sample described in
\S~\ref{subsec:galaxysample} according to the methods discussed in
\citet{berlind06}. Briefly, groups are identified via a redshift-space
friends-of-friends algorithm that takes no account of member galaxy
properties beyond their redshifts and positions on the sky. For each
group, we define the central galaxy to be the brightest
non-fiber-collided member\footnote{See \citet{skibba11} for a discussion of the complicating factor that central galaxies are not always the brightest group members.}, and the group's satellites to be the
remaining non-fiber-collided members. The details of how
fiber collisions influence central/satellite designations in this
group sample was studied extensively in \citet[][hereafter
  H12]{hearin_etal12b}, and we follow their conventions here. We refer
the reader to \citet{berlind06} for further details on the group
finding algorithm.


\begin{figure*}
\begin{center}
\includegraphics[width=.9\textwidth]{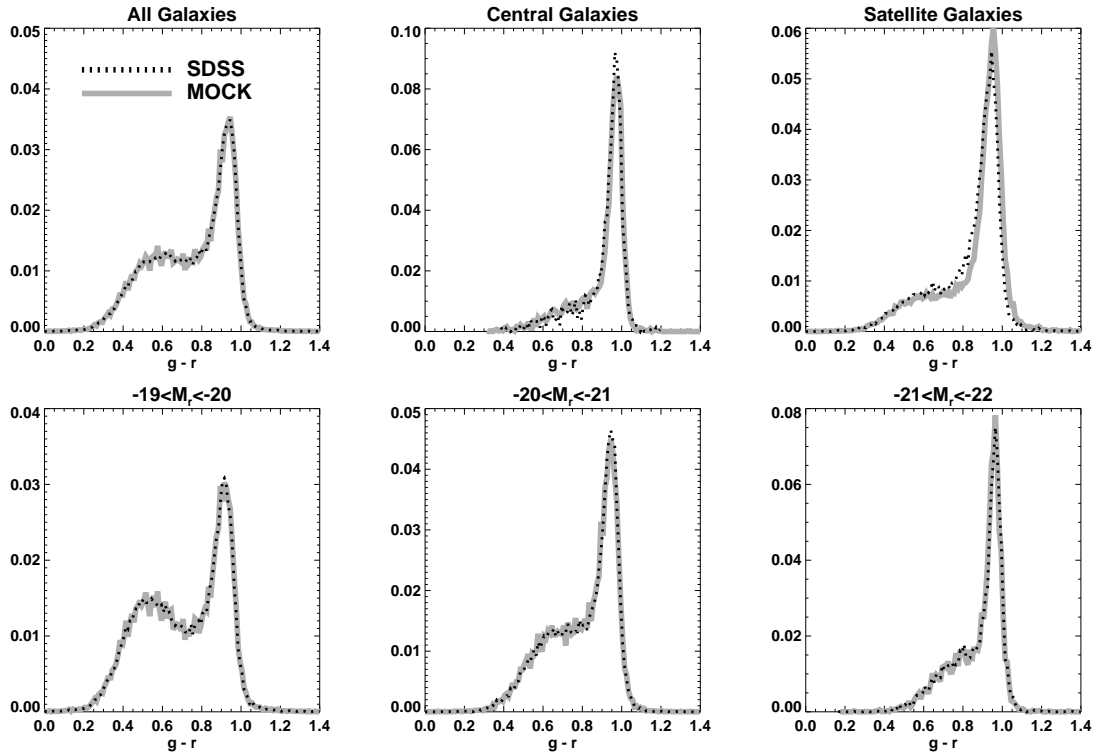}
\caption{Galaxy color PDFs from our mock catalog as compared to those
  measured in the Mr19 SDSS galaxy catalog.  By construction, the $(g-r)$ probability distribution of our mock galaxies 
  is in exact agreement with the
  data (black dotted histograms) at all luminosities. This is evident in the top left panel, and in all three bottom panels. Additionally, 
  we compare color PDFs for central and satellite galaxies in the group catalog in the top center and top right panels, respectively. 
  Since information about satellite/central designation does not inform
  the colors we assign to the galaxies, the center and right panels on the top row illustrate successful {\em predictions} of distribution age matching.}
\label{fig:mock_PDFs}
\end{center}
\end{figure*}



\section{Simulation and Halo Catalogs}
\label{sec:sims}


Throughout this work we use the high resolution, collisionless, cold
dark matter $N-$body Bolshoi simulation \citep{bolshoi_11}.  The
simulation has a volume of $250^{3}\,\hmpcvol$ with $2048^{3}$
particles and a cold dark matter ($\Lambda$CDM) cosmological
model with $\Omega_{\mathrm{m}}=0.27$, $\Omega_{\Lambda}=0.73$,
$\Omega_{\mathrm{b}}=0.042$, $h=0.7$, $\sigma_{8}=0.82$, and
$n_{\mathrm{s}}=0.95$.  Bolshoi has a mass resolution of $1.9\times
10^8\,\Msun$ and force resolution of $1\,\hkpc$, and particles were
tracked from $z=80$ to $z=0$.  Bolshoi was run with the Adaptive
Refinement Tree Code (ART; \citealt{kravtsovART97, gottloeber_klypin08}).
The Bolshoi data are available at {\tt http://www.multidark.org} and
we refer the  reader to \citet{riebe_etal11} for additional
information.

Halos and subhalos were identified with the phase-space temporal halo
finder ROCKSTAR \citep{rockstar_trees,rockstar}, capable of resolving Bolshoi
halos and subhalos down to 
$\vmax\sim55\kms$.   Halo masses were calculated using spherical
overdensities according to the redshift-dependent virial overdensity
formalism of \citet{bryan_norman98}.  We use the $z=0$ ROCKSTAR halo
catalog and merger trees to model the galaxy color-halo connection.


\section{Mock Catalogs}
\label{sec:mocks}


In this section we describe our algorithm for constructing our mock
sample of galaxies and galaxy groups. We apply { age distribution matching}, defined in \S~\ref{subsec:soam}, to the
ROCKSTAR halo catalogs at $z=0$ to construct a sample of mock galaxies
with both $r$-band luminosities $M_r$ and $(g-r)$ colors. After
constructing the mock galaxy sample from the halo catalog, we place
the mock galaxies into redshift space and apply the same group-finder
to the mock data that was applied to the SDSS Mr19 galaxy sample (see
\S~\ref{subsec:sdssgroups}). We now discuss the steps of this
procedure in turn. 


\begin{figure*}
\begin{center}
\includegraphics[width=.9\textwidth]{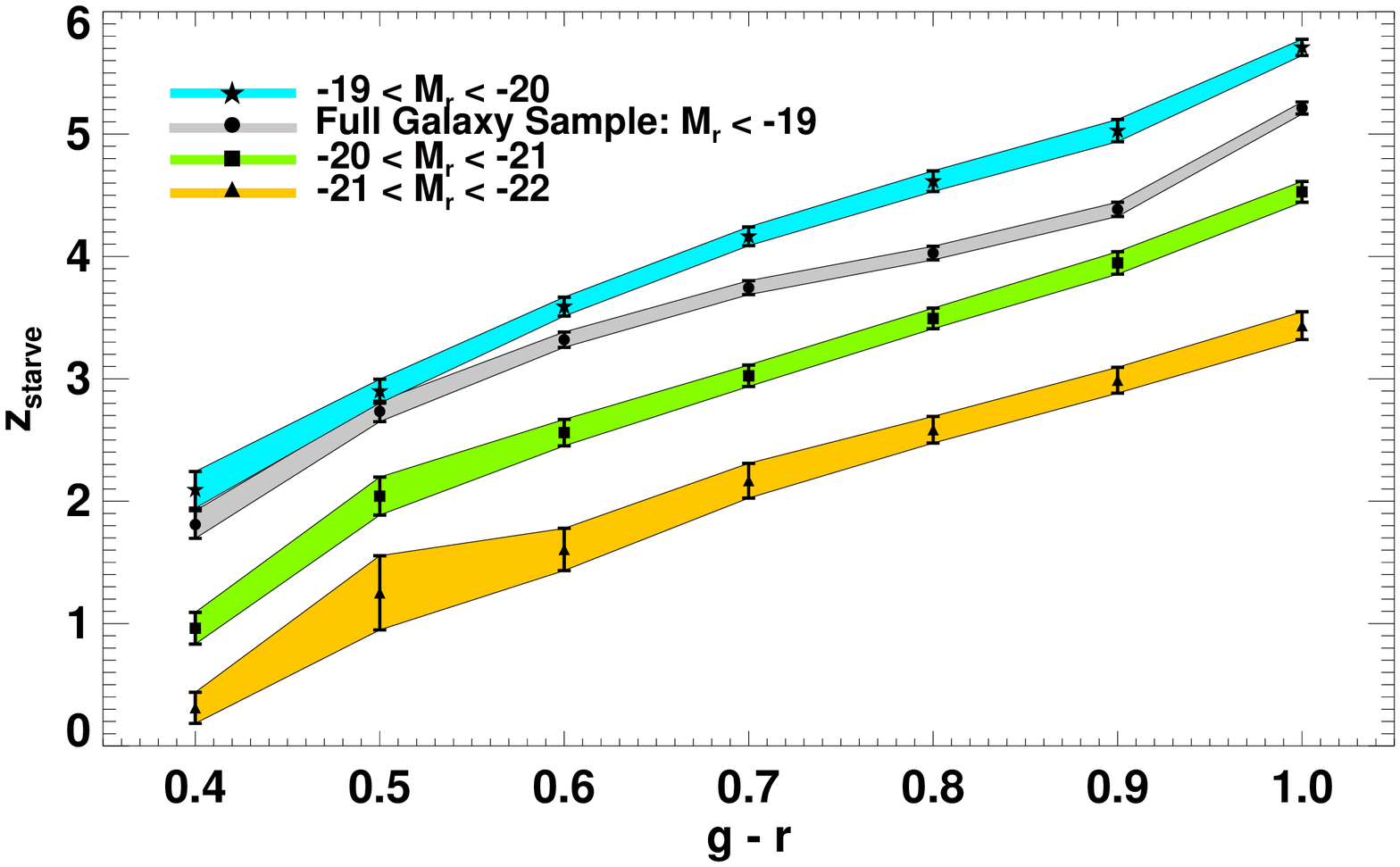}
\caption{
Implied relation between $g-r$ color and $\zstarve.$ Results for the $M_{r}<-19$ threshold sample appear in gray, luminosity-binned samples are coded according to the legend. Error bars are estimated by bootstrapping. The relation varies between luminosity bins due to the luminosity-dependence to the color PDF $\Psdss.$ All results demonstrate the defining assumption of age distribution matching, that  galaxies with older stellar populations are typically found in older halos.}
\label{fig:zstarvegr}
\end{center}
\end{figure*}


\subsection{Age Distribution Matching}
\label{subsec:soam}

In this section we describe our two-phase algorithm for assigning
both luminosity and color to the galaxies located at the center of
(sub)halos in an N-body simulation. We describe the luminosity assignment phase in  \S~\ref{subsubsec:mrassign}, the color assignment phase in \S~\ref{subsubsec:grassign}, and provide a summary in \S~\ref{subsubsec:admbullets}.

\subsubsection{Luminosity Assignment}
\label{subsubsec:mrassign}

We assign $r$-band luminosity to Bolshoi (sub)halos using the standard
abundance matching technique. This technique has been widely used in
the literature and so we do not discuss all of the technical details
of this phase of our algorithm. Our implementation of the
abundance matching luminosity assignment is identical to that
described in detail in Appendix A of H12, and thus we restrict
ourselves to an overview here.

The quantity $\vmax\equiv \mathrm{Max}\left\{ \sqrt{GM(<r) / r}
\right\},$ where $M(<r)$ is the mass enclosed within a distance $r$ of
the halo center, defines the maximum circular velocity of a test particle orbiting in the halo's gravitational
potential well. The abundance matching algorithm
for luminosity assignment assumes a monotonic relationship between
luminosity and $\vmax,$ such that the cumulative abundance of SDSS
galaxies brighter than luminosity $L,$ $n_{g}(>L),$ is equal to the
cumulative abundance of halos with circular velocity  larger
than $\vmax,$ $n_{h}(>\vmax).$   Following common practice we abundance match on the property $\vpeak,$ the maximum value $\vmax$ ever
attains over the entire merger history of the (sub)halo
\citep[see][for a thorough discussion of the use of $\vpeak$ in
  abundance matching]{reddick12}. 

Recent results \citep{wetzel_white10} indicate that halos at the faint end of our mock galaxy sample ($M_{r}\lesssim-19$) may be pushing the resolution limit of 
the Bolshoi simulation. We do not attempt to correct for possible resolution effects, but note that such corrections \citep[e.g.,][]{moster10} may be important for precision predictions of the small-scale 2PCF. A systematic investigation of these resolution effects is beyond the scope of this paper, and so we leave this as a task for future work. 


\begin{figure*}
\begin{center}
\includegraphics[width=1.\textwidth]{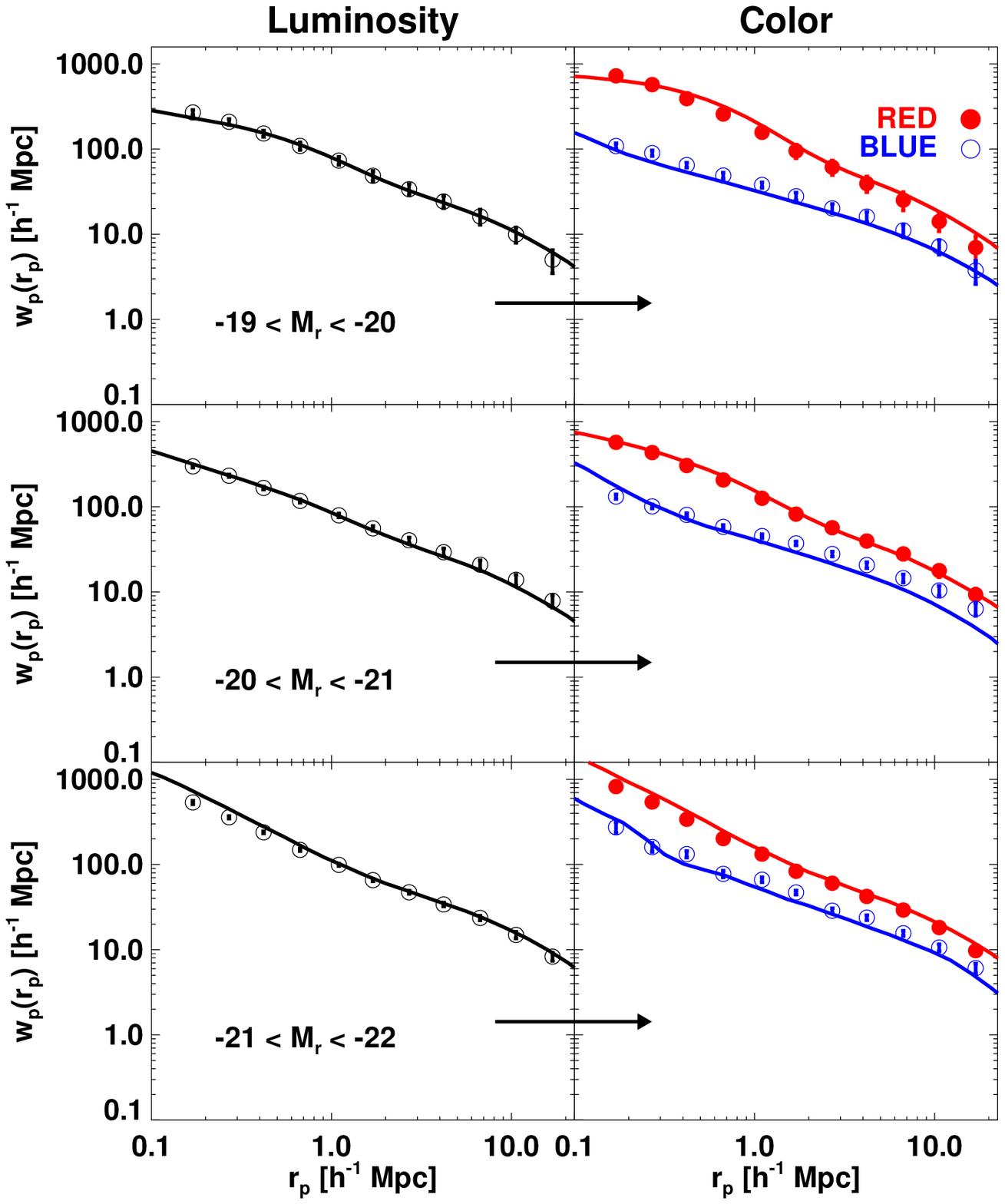}
\caption{Luminosity- and color-dependent clustering as predicted by
  our age distribution matching formalism. \emph{Left Column}:
  The luminosity-binned projected 2PCF predicted by our model (black
  solid curves)  against the clustering exhibited by SDSS galaxies.
  \emph{Right Column}: In bins of luminosity, we plot the projected 2PCF of red
  (blue) mock galaxies with red (blue) solid curves.  Red, filled
  (blue, open) points show the clustering of red (blue) SDSS
  galaxies. We use the results presented  in \citet{zehavi11} for all
  SDSS measurements shown in this figure.}
\label{fig:lum_color_wp}
\end{center}
\end{figure*}

 
Galaxy formation is a complex process, and so a single halo property
such as $\vmax$ cannot uniquely specify the luminosity of the halo's
galaxy. For example, the Tully-Fisher relation has intrinsic scatter
\citep{pizagno07}, and so abundance matching cannot reproduce the
statistics of the observed galaxy distribution without scatter
\citep[see][for a comprehensive analysis of sources of scatter in
  abundance matching modeling]{behroozi10}. Our luminosity abundance
matching technique results in $\sim0.15$dex of scatter in luminosity at fixed $\vmax$ over the
luminosity range of our sample, in accord with results from satellite kinematics \citep{more09b} as well as other abundance matching studies \citep{reddick12}.  Details of our implementation of
scatter can be found in Appendix A of H12.

\subsubsection{Color Assignment}
\label{subsubsec:grassign}

Once luminosities have been assigned to the mock galaxies, we proceed
to the second phase of our algorithm. First, we divide the mock and
SDSS galaxies into ten evenly spaced bins of $r$-band magnitude. We
use the observed colors of the SDSS Mr19 galaxies in each luminosity
bin to define the probability distribution $\Psdss.$ If there are
$N_{\mathrm{L}}$ mock galaxies in luminosity bin $L,$ we randomly draw
$N_{\mathrm{L}}$ times from $\Psdss$ and rank-order the draws, reddest
first. We assign these colors to the mock galaxies in luminosity bin
$L$ after first rank-ordering these $N_{\mathrm{L}}$ mock galaxies by
the property $\zstarve$ (defined below), largest first. 

By repeating the above procedure in each luminosity bin, the color
distribution of the mock galaxies is in exact agreement with the data,
as can be seen in the top left and bottom panels of Figure~\ref{fig:mock_PDFs}. Note that this desirable
property of our mock would be true \emph{regardless} of the rank-ordering
on $\zstarve.$ The effect of the rank-ordering is to introduce, at
fixed luminosity, a correlation between galaxy color and the epoch of starvation.

The property $\zstarve$ is based on three characteristic epochs of a
halo's mass assembly history.  These include: \ben
\item[\textbf{1.}]{ $\zchar:$ The first epoch at which halo mass
  exceeds $10^{12}\hMsun.$ For halos that never attain this mass
  $\zchar=0.$}
\item[\textbf{2.}]{$\zacc:$  For subhalos,
  $\zacc$ is the epoch after which the object always remains a
  subhalo. For host halos, $\zacc=0.$}
\item[\textbf{3}]{$\zform:$ Using the methods of \citet{wechsler02} we
  identify the  redshift at which the halo transitions from the fast-
  to slow-accretion regime, as this is the epoch after which  dark
  matter ceases to accrete onto the halo's central region.}  \een From
  these three characteristic epochs we define the redshift of 
  starvation:  \beq
\label{eq:zstarvedef}
\zstarve\equiv \mathrm{Max}\left\{\zacc,\zchar,\zform\right\}.   \eeq

We provide a full discussion of the physical motivation and
interpretation of $\zstarve$ in \S~\ref{subsec:zstarvephysics}. In Appendix
A, we discuss the details of how each of the above properties are
computed from Bolshoi merger trees. 

\subsubsection{Algorithm Summary}
\label{subsubsec:admbullets}

For convenience, we conclude discussion of our mock-making algorithm by briefly summarizing the steps discussed above:
\ben
\item From the Bolshoi halo catalog and merger trees, we compute $\vpeak$ and $\zstarve$ for every (sub)halo in the simulation.
\item We assign r-band luminosities to the (sub)halos using abundance matching with the halo property $\vpeak.$ Mock galaxies dimmer than $M_{r}>-19$ are cut from the sample.
\item The mock galaxies are binned by luminosity, and colors are assigned by randomly drawing from the color PDF $\Psdss$ defined by our SDSS Mr19 galaxy sample. The random draws from $\Psdss$ are rank-ordered, and the mock galaxies in each luminosity bin are rank-ordered by the halo property $\zstarve,$ such that in each luminosity bin mock galaxies where $\zstarve$ occurs earlier are assigned redder colors. 
\een

\subsection{Group Identification}
\label{subsec:mockgroups}

Once colors and luminosities have been assigned to the halo catalog,
we place our mock galaxies into redshift space and run the same group-finding
algorithm on the mock that was run on the SDSS Mr19 galaxy sample. In
this way, our mock groups inherit the same purity and incompleteness
systematics as our SDSS groups. Finally, we introduce fiber collisions
into our mock sample according to the prescription adopted in H12;
this method preserves the relationship between local galaxy density
and the likelihood that a fiber collision occurs in the SDSS Mr19
sample.



\section{Results}
\label{sec:results}


We now investigate how well our model can predict a variety of observed galaxy
statistics.  First, we illustrate the general relationship between $(g-r)$ and $\zstarve,$ and then demonstrate that our model naturally predicts the correct relative colors of central and satellite galaxies.
We show the accuracy of our base Mr19 SDSS
catalog at reproducing the observed SDSS luminosity-binned 2PCF,
explicitly showing that our mock catalog naturally inherits the
successes of  abundance matching.  We then compare the
resulting color-binned 2PCF predictions to SDSS measurements.
Finally, we test the success of our model against a variety of galaxy
group statistics measured from the Mr19 group catalog.


\begin{figure*}
\begin{center}
\includegraphics[width=1.\textwidth]{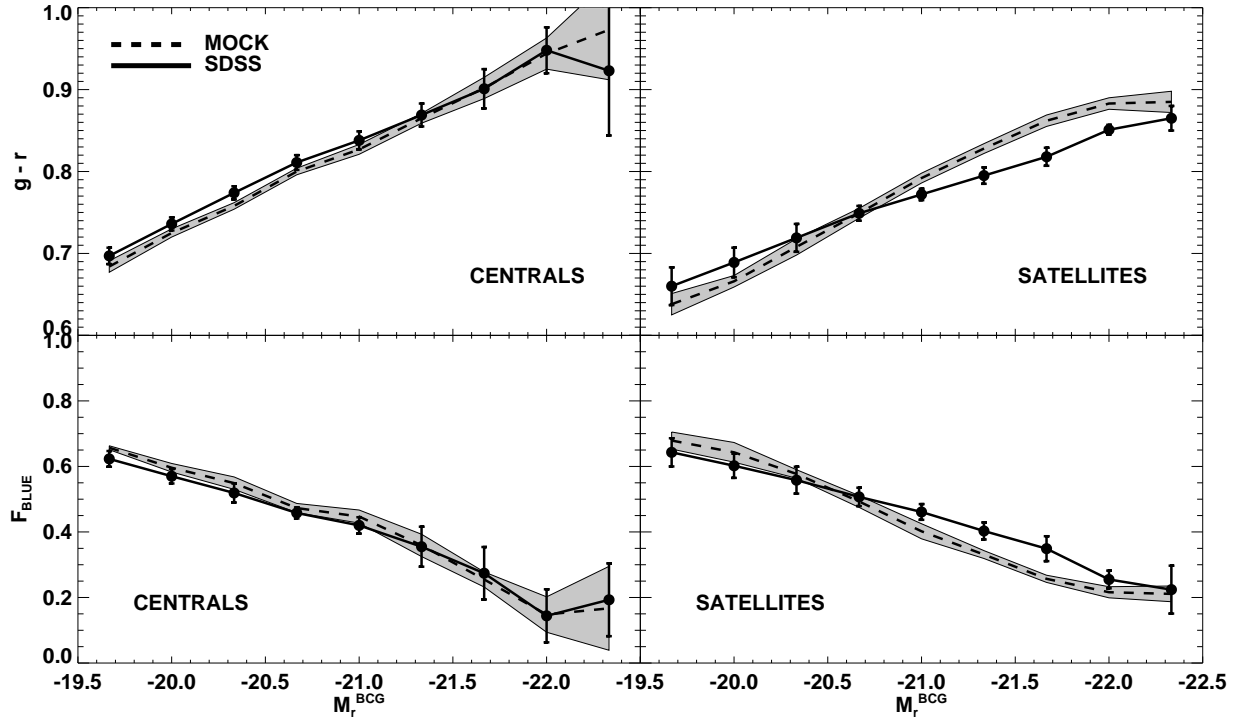}
\caption{\emph{Top Row}: Mean $(g-r)$ color as a function of absolute
  magnitude of the brightest central galaxy,
  $M_{r}^{\mathrm{BCG}}$ for  central galaxies (left panel)
  and satellite galaxies (right panel).  SDSS data  are shown as the
  black filled circles and the dashed curves represent the predictions
  from our mock catalog.  Solid gray bands are the errors computed as
  the dispersion over 1000 bootstrap realizations of the galaxy sample
  (see \S~\ref{subsec:groups_results} for details). \emph{Bottom Row}:
  Same as the top row but now with the fraction of galaxies in groups
  of a given  $M_{r}^{\mathrm{BCG}}$ which are blue,
  $\mathrm{F_{BLUE}}$, as defined by the color cut in
  Eq.~\ref{eq:colorcut}.  }
\label{fig:color_fblue_Lbcg}
\end{center}
\end{figure*}


\subsection{Halo Age and Galaxy Color}

In Figure \ref{fig:zstarvegr}, we plot the mean $\zstarve$ value of (sub)halos in our mock in bins of $(g-r)$ color. The gray band shows the result for all mock galaxies in our sample, while the result in bins of luminosity appear in the colored bands. The width of the bands corresponds to bootstrap estimates of the error on the mean. This figure visually demonstrates the fundamental assumption of age distribution matching: redder galaxies tend to live in older halos. The luminosity dependence of the $\zstarve-$color relation simply reflects the color-luminosity trend seen in the data, $\Psdss.$ 

The values of $\zstarve$ may seem strikingly large. Indeed, the reddest galaxies in our model typically have $\zstarve\approx5-6.$ This highlights an important point: {\em the property $\zstarve$ does not correspond to the redshift where star formation in the galaxy is quenched.} Rather, $\zstarve$ signifies a special epoch in halo assembly that correlates strongly with the epoch when star formation in the galaxy becomes inefficient. In our color assignment algorithm (described in \S~\ref{subsubsec:grassign}), rank-ordering on $\zstarve$ only serves to govern the {\em relative} colors assigned to halos; drawing the assigned colors from $\Psdss$ ensures that the absolute values of the assigned colors are correct. Thus in the (likely) event that there is a substantial time delay between $\zstarve$ and $\zquench,$ provided that this time delay does not strongly vary with halo mass then rank-ordering on $\zstarve$ should still suffice to assign the right colors to the right halos. We return to this point in \S~\ref{sec:discussion}.

\subsection{Central and Satellite Colors}

As discussed in \S~\ref{subsubsec:grassign}, our model correctly reproduces the color distribution $\Psdss$ by construction. 
However, this by no means guarantees that our color PDF will be correctly predicted when conditioned on some other property besides luminosity. 
In the top center and top right panels of Figure~\ref{fig:mock_PDFs}, we show the color PDFs of our entire sample, conditioned on whether the galaxy is a central or satellite, respectively. 

Our color assignment only uses the property $\zstarve$ and $\Psdss$ to assign colors to the mock galaxies but does not  distinguish between centrals and satellites. Thus the good agreement seen between our model and the data in the top center and top right panels of Figure~\ref{fig:mock_PDFs} demonstrates a successful {\em prediction} of age distribution matching. In our model, central and satellite galaxies have different color distributions because host halos and subhalos have different mass assembly histories.

\subsection{Luminosity- and Color-Dependent Clustering}\label{subsec:2PCF_results}

We begin discussion of our clustering results by showing that our abundance matching-based luminosity
assignment (described in \S~\ref{subsubsec:mrassign}) accurately
predicts the observed luminosity-binned 2PCF as measured by
\citet{zehavi11}.  To do this, we take our volume-limited Mr19 catalog
and sub-divide into three luminosity bins: $-19 \leq M_{r}  \leq -20$,
$-20 \leq M_{r}  \leq -21$, and  $-21 \leq M_{r} \leq -22$.  For each
of these luminosity bins, we compute the real-space 2PCF.  We then
convert to projected space, $\wprp$, via:
\begin{equation}
  \wpp\rp =
  2\int_0^{\pimax}\xi\Big(\sqrt{r_{\mathrm{p}}^2+\pi^2}\Big)d\pi. 
\end{equation}
For the upper limit in the integration, we use $\pimax=40$ $h^{-1}\mathrm{Mpc}$ to be consistent with the measurements made for our observational samples.

The black solid curves in the left column of
Figure~\ref{fig:lum_color_wp} show the 2PCF as predicted by our mock
catalog.  There is excellent agreement with the \citet{zehavi11} SDSS
measurements (black data points) at each luminosity bin and at all
projected separations from the linear to highly non-linear regimes.

From this success, we now turn to the right column of
Figure~\ref{fig:lum_color_wp} showing the color-dependent 2PCFs  in
distinct $r$-band luminosity bins as predicted by our age distribution matching formalism.  Red filled and blue open circles in all
panels represent the red and blue galaxy populations from
\citet{zehavi11}, respectively.  Red and blue solid curves are the
model predictions.  While there is some discrepancy between the clustering of blue galaxies in the 
model and the data at large scales for the $-20 \leq M_{r}  \leq -21$
bin, overall there is good agreement for each luminosity bin
and at all $r_p$ separation scales.  This result is very encouraging
since the color-dependent 2PCF encodes rich information about the galaxy-halo connection \citep[e.g.,][]{more_etal13,cacciato_etal13}. However, as shown in \citet{masaki13} the 2PCF alone is not enough to discriminate between competing models for the relationship between a halo and the color of the galaxy it hosts (see also \citet{neistein11}). We address this in the next section with our exploration of galaxy group statistics. 

We emphasize that we have done no parameter fitting to achieve the
agreement between the predicted and measured color-binned
clustering. Our algorithm for color assignment has no explicit
dependence on halo position; rather, the clustering signal in our mock
emerges as a {\em prediction} of our theory. Since the primary aim of
this paper is to introduce the age distribution matching
formalism, we have opted for the simplest implementation of this
framework rather than attempting to finely tune our model according to
the measured clustering (see \S~\ref{sec:discussion} for a discussion
of more complex possible implementations). The success of our model's
prediction for the color-binned 2PCF is rather striking given
the simplicity of the formulation we present here.

\subsection{Galaxy Group Statistics}\label{subsec:groups_results}

In this section we test how well our model distinctly assigns colors
to satellite and central galaxies in our mock catalog compared to
those colors in the SDSS galaxy group catalog. In the {\em top row} of
Figure~\ref{fig:color_fblue_Lbcg}, we show the mean  $(g-r)$ color of
group galaxies as a function of the brightness of the group's central
galaxy, $M_{r}^{\mathrm{BCG}}$.   From left to right, we show
the results for central galaxies and satellites, respectively, in the
mock and the SDSS group catalog.  The dashed line traces the mean
$(g-r)$ color of the mock galaxies in a given
$M_{r}^{\mathrm{BCG}}$ bin.  The {\em bottom row} is similar
to the top, except we plot $\fblue,$ the fraction of
galaxies that are designated `blue' according to the  color cut
defined by Eq.~\ref{eq:colorcut}. 

In all panels of Fig.~\ref{fig:color_fblue_Lbcg}, the solid gray
region shows the errors on the mean.  We use standard bootstrap
resampling techniques to estimate the errors in all of our group-based
statistics. We compute the errors as the dispersion over 1000
bootstrap realizations of the galaxy sample, where each realization is
constructed by randomly selecting\footnote{We bootstrap resample {\em
    with replacement}, that is, we allow for the possibility of
  repeated draws of the same object.} $N_g$ galaxies from the sample,
where $N_g$ is the total number of galaxies in the sample.  

The successful, distinct predictions for the colors of central and
satellite galaxies emerges naturally from the age distribution matching formalism. At fixed luminosity, the colors of our mock
galaxies are drawn from the same color PDF, $\Psdss,$ regardless of
sub/host halo designation. Moreover, our color assignment algorithm
takes no explicit account of (sub)halo mass. Thus the difference
between the colors of satellites and centrals in our mock is purely a
reflection of the different mass assembly history of host halos and
subhalos. Excepting only some mild tension for the color of satellite galaxies in a few of the brighter $M_{r}^{\mathrm{BCG}}$ bins,  
the agreement between mock and
observed central and satellite galaxy color is quite good. Again, we emphasize that we have not tuned any parameters in our model and
we have chosen to keep our implementation of the age distribution matching technique as simple as possible (see
\S~\ref{sec:discussion} for an elaboration of this point). 


\section{DISCUSSION \& IMPLICATIONS}\label{sec:discussion}


\subsection{Comparison to Abundance Matching}

Mock catalogs based on age distribution matching inherit
all the successes of abundance matching because this is precisely the
method we use in the luminosity assignment phase of our algorithm. The
color PDF $\Psdss$ plays the same role in the color assignment that
the luminosity function $\Phi_{\mathrm{SDSS}}(L)$ plays in the
luminosity assignment. Thus just as $\Phi_{\mathrm{mock}}(L)$ from
  abundance matching agrees precisely with the observed
luminosity function, so do our mock color distributions agree
precisely with the observed color PDF. 

There is an important conceptual difference between abundance matching  and
age distribution matching: the role that the property $\zstarve$
plays in the color assignment differs markedly from the role played by
$\vmax$ in the luminosity assignment. In   abundance matching, luminosity assignment simply
cannot proceed without appeal to the property $\vmax$ (or
  some other halo property such as $M_{\mathrm{vir}}$ for host halos
  and $M_{\mathrm{peak}}$ for subhalos),  because this is an
essential ingredient to the defining relation $n_{\mathrm{g}}(>L) =
n_{\mathrm{h}}(>\vmax).$ This is not the case with the color
assignment. While the dark matter halo property $\zstarve$ is in monotonic
correspondence with $(g-r)$ color just as $\vmax$ is in monotonic
correspondence with luminosity, this feature is not essential in order
for the color assignment algorithm to proceed. Indeed, the
rank-ordering of $(g-r)$ color with $\zstarve$ only serves to
correlate galaxy color with halo mass assembly history (at fixed $\vmax$), but a mock
catalog constructed without this correlation would nonetheless have a
color PDF that is in agreement with $\Psdss.$

In light of this observation, one may wonder whether correlating
galaxy color with $\zstarve$ is necessary at all in the construction
of a successful model. However, a mock catalog constructed by drawing colors from the data without any rank-ordering predicts no difference 
between the clustering of red and blue samples (at fixed luminosity). This failure is easy to understand: in such a model colors are assigned purely at random, and so there is no mechanism which produces any trend between color and the cosmic density field.

\subsection{Comparison to HOD models}

In the standard implementation of the HOD formalism, the only halo property that governs galaxy occupation is the host halo mass, $\mhost.$ In particular, 
both \citet{zehavi11} and \citet{skibba09} employ the HOD in their distinct approaches to modeling color-dependent clustering, and in neither model is there any explicit use of halo age. In the \citet{zehavi11} model, the HOD parameters of red and blue galaxy samples are constrained by two-point clustering together with galaxy abundance. In the alternative HOD approach explored in \citet{skibba09}, the authors draw $(g-r)$ colors for mock centrals and satellites from explicitly different distributions. Both models faithfully reproduce the color-dependence of the 2PCF (see also \citet{skibba13} for tests of color HOD models beyond binned 2PCF statistics). 

The above HOD-based models assume that there are only two pieces of information that determine (in a statistical sense) the color of a galaxy: 1) whether the galaxy is a central or a satellite, and 2) the mass of the host halo in which the galaxy resides. As both of these models enjoy great success in reproducing observations, it may seem surprising that our model also succeeds since the color assignment phase of our algorithm appears to take no account of host halo mass. Moreover, in our model the only distinction between centrals and satellites is the use of $\zacc,$ yet $\zstarve\neq\zacc$ for the majority ($>90\%$) of subhalos at all luminosities. Thus our model also appears to assign colors to central and satellite galaxies in an identical fashion, at least in effect.

The resolution to this puzzle is illuminating. First, in CDM there is a strong correlation between halo mass and formation, with more massive halos forming at later times. Thus our model effectively {\em does} introduce rather strong trends between $(g-r)$ color and host halo mass due to the halo age-mass correlation. Thinking of $M_{r}^{\mathrm{BCG}}$ as a proxy for host halo mass, these trends can readily be seen in Figure~\ref{fig:color_fblue_Lbcg}. 

Second, at fixed mass, satellites and centrals have quite different assembly histories. This is true even when accounting for post-accretion stripping by comparing centrals to satellites at fixed mass at accretion, $M_{\mathrm{acc}}.$ In general, at fixed $M_{\mathrm{acc}},$ satellites formed earlier than centrals, since satellites formed in denser regions of the initial density field and rapidly accreted their mass at earlier times. In age distribution matching, it therefore naturally emerges that satellites and centrals have different colors (Figure~\ref{fig:mock_PDFs}) even though we assign colors to them from the same PDF.

We conclude our discussion of models based only on host halo mass  by noting that any model for galaxy color that depends strictly on $\mhost$ is unable even in principle 
to reproduce the phenomenon of so-called {\em Galactic Conformity}: at fixed host mass, group systems with a blue central have a bluer satellite population than groups with a red central \citep{weinmann06b}. 
This feature emerges naturally in the age distribution matching formalism, as we will demonstrate explicitly in a future companion paper. 

\subsection{Physical Motivation for $\zstarve$}
\label{subsec:zstarvephysics}

The expectation that $\zstarve$ should be related to stellar mass assembly
is physically well-motivated. First, it is natural to expect that $\zform$ represents an important
epoch in the history of the formation of a galaxy. As shown in
\citet{wechsler02}, at $\zform$ the dark matter halo transitions from
the fast- to slow-accretion regime. After this epoch the matter
accreting onto the dark matter halo ceases to penetrate into the core
of the halo where the galaxy resides\footnote{See their Figure 18 for
  a demonstration of this point.}, depriving the galaxy of new material from the
field with which it can form stars. 

In reality, the halo may continue to form stars from its existing gas supply after this epoch, leading to a delay 
between $\zform$ and the actual quenching of star formation. 
However, our model is based on {\em rank-ordering} of the property $\zstarve,$ and so the {\em relative} color of galaxies in halos with different 
formation times will be correct provided that this time delay is not a strong function of redshift or halo mass. 
As discussed in \S~\ref{subsec:future}, we will explore more complex models with quenching time delays in a future companion paper. 
We simply note here that corrections due to mass- and redshift-dependent time delays cannot be too severe or our model would not successfully 
predict the 2PCF and group-based statistics. 

Second, the existence of a characteristic halo mass $\mchar$ above 
which star formation is highly inefficient also has strong empirical support. 
The halo mass-dependence of the stellar mass to halo mass ratio of galaxies shows a peak at $\mchar\sim10^{12}\hMsun$ 
that rapidly falls off as halo mass increases \citep{yang12,yang13,watson_conroy13,behroozi12,moster13}. This characteristic mass was recently shown to remain essentially 
constant for most of cosmic history \citep{behroozi_etal13b}. 
Additionally, $\mchar\sim10^{12}\hMsun$ is the halo mass at which pressure-supported shocks can heat infalling gas to the virial temperature \citep{dekel_birnboim06}. 
More importantly, this is also the mass when effects due to AGN feedback become significant, which can have 
a dramatic effect on star formation \citep{shankar_etal06,teyssier11,martizzi_etal12}. 

In our implementation of age distribution matching, we frame no
hypothesis for the particular physical mechanism(s) that influence
star formation inside massive halos. Instead, we simply posit that there exists a
characteristic halo mass $\mchar$ above which star formation becomes
inefficient, and that this leads to a correlation between galaxy color
and the characteristic epoch $\zchar$ that the halo attains this
mass. 

We explored a range of masses for $\mchar$ and find that both
the 2PCF and our group statistics allow us to discriminate between
competing values for the characteristic mass. In models where
$\mchar\gtrsim10^{12.5}\hMsun,$ bright central galaxies are far too
blue, and bright blue galaxies are strongly over-clustered on small
scales ($\lesssim1\hmpc$). Conversely, in models with
$\mchar\lesssim10^{11.5}\hMsun$ central galaxies are too red and the
2PCF is in poor agreement with the data in all luminosity bins. 

Finally, we turn to the property $\zacc,$ the epoch a satellite
accreted onto its host halo. Numerous studies have shown that the
dependence of galaxy color and/or star formation rate (SFR) has
little-to-no residual environmental dependence once small-scale
environment ($\lesssim1\hmpc$) has been accounted for
\citep[e.g.,][]{hogg04,blanton05}. This small-scale environmental
dependence has been shown to be driven primarily by correlations
between the properties of the host halo and its satellite galaxies
\citep{blanton_berlind07,wetzel_etal11,tinker_wetzel11,wetzel_etal12b}. These
results have given rise to a widely accepted picture in which satellite evolution is significantly influenced by 
processes that occur near or within the virial radius of the host halo.

There are a variety of physical mechanisms that may have a significant impact on
satellite galaxy color/SFR. These include, but are not limited to, 
mass loss due to outside-in stripping \citep[see][and
  references therein]{watson_etal12b}, 
  ``strangulation" \citep{larson80,diemand_etal07,kawata_mulchaey08}, 
  ram-pressure stripping \citep{gunn_gott72,abadi_etal99,mccarthy_etal08},
  ``harassment" \citep{moore_etal98}, and intra-host mergers
\citep{makino_hut97,wetzel_etal09b,wetzel_etal09a}. 
Again, in our implementation of age distribution matching, we do
not preferentially select any one of these physical mechanisms as the
primary cause for satellite quenching. Instead, we make the following
simple observation: the impact that each of the scenarios discussed
above has on satellite color/SFR will be greater for satellites that
fell into their host halo potential at earlier times. Thus it is
natural to expect a correlation between satellite galaxy color and
$\zacc$. Accordingly, we identify $\zacc$ as the third 'special' epoch in
(sub)halo mass accretion history.

Our treatment of the effect satellite accretion has on color/SFR is consistent 
with the finding in \citet{wetzel_etal11} that there is no minimum host halo mass for post-accretion physics to significantly influence satellite evolution. 
Moreover, since we correlate color with $\zacc$ regardless of host halo mass, the success of our implementation of age distribution matching 
provides new supporting evidence for conclusions drawn in \citet{vdBosch08}, who found that 
satellite-specific processes are equally efficient in host halos of all masses.

With the above physical picture in mind, in age distribution matching we make the following simple assumption: {\em the earlier any
  one of the above processes begins in a halo, the redder its galaxy
  will be today.} Mathematically, this is formulated by introducing a
monotonic correspondence between $(g-r)$ color and
$\zstarve\equiv\mathrm{Max}\left\{\zform,\zchar,\zacc\right\},$ as
described in \S~\ref{subsubsec:grassign}.

In alternative models that neglect to account for $\zchar$ and instead define $\zstarve\equiv\mathrm{Max}\left\{\zform,\zacc\right\},$ the clustering prediction for bright galaxies is in very poor agreement with the data. It is natural to expect that this failure would occur at the bright end: in such a model, group/cluster BCGs are all bright blue! This is because the halos hosting these BCGs are very massive and therefore very late-forming, and so without $\zchar$ these halos have the smallest values of $\zstarve$ in the halo catalog. Conversely, models that neglect to account $\zform$ perform well at the bright end but fail to correctly predict the clustering of dimmer galaxies. Thus both $\zchar$ and $\zform$ are essential ingredients to the success of age distribution matching. 

The same is not true for $\zacc:$ models which do not account for subhalo accretion make quite similar predictions to those that do. This is because subhalos are typically not very massive objects that form very early on, and so for the typical subhalo $\zstarve\gg\zacc.$ This result may seem surprising since so much of the literature on star formation focuses squarely on the influence of satellite-specific quenching. Thus the success of our current model suggests that too much importance may have been placed on post-accretion physics \citep[see also][for an alternative manifestation of this point]{watson_conroy13}. However, since there is likely a long time delay between $\zstarve$ and the actual quenching of star formation, we anticipate that subhalo accretion will prove to be more important when we more thoroughly explore models based on $\zquench\equiv\zstarve+\Delta t_{\mathrm{delay}}.$ We return to this point in \S~\ref{subsec:future}.

\subsection{Comparison to other phenomenological models}

In a recent paper closely related to this work, \citet{masaki13} paint
a binary `red' or `blue' color onto (sub)halos that have first been
assigned luminosities based on   abundance matching. The authors explored a
variety of different halo properties to use in the color assignment. They found that several different choices result in successful
predictions for the 2PCF, motivating their use of galaxy-galaxy
lensing measurements to discriminate between competing models. Since
precision galaxy-galaxy lensing measurements that have been binned simultaneously on both r-band luminosity and 
$(g-r)$ color were not available, the authors instead used measurements
binned on morphology and assumed a perfect correspondence between red
(blue) galaxies and early (late) type morphological classification.

While the assumption that all red (blue) galaxies have early (late) type
morphology is a reasonable first approximation \citep{sheldon04}, we do not rely on it in
this work. Instead, we employed galaxy group-based statistics to provide our
additional discriminating power. Galaxy group membership yields
information about halo occupation that is independent from 2PCF
information. Group membership is necessarily identified in redshift
space, and so group-based statistics incorporate information based on
the velocity field, whereas projected clustering does
not. Additionally, group statistics can probe halo occupation at
higher masses than it is possible to do with small-scale clustering
\citep[see][Figure 2]{hearin_etal12}. Nonetheless, in a future
companion paper to this one, we will make new galaxy-galaxy lensing
measurements of SDSS galaxies  binned on $(g-r)$ color to provide
further tests of the success of the model we present here.

The semi-analytic modeling (SAM) approach to predicting galaxy
properties such as color and star formation differs markedly from the
theory presented here. SAMs also use the hierarchical build-up of dark
matter halos as a foundation, but from here SAMs model the myriad  baryonic processes
that take place inside halos with a set of analytic formulae together
with a  large number of finely tuned parameters
\citep{white_frenk91,kauffmann_etal99,cole_etal94,somerville_primack99,DeLucia_Blaizot07,guo_etal11b}. Such approaches have been
quite successful  at reproducing many statistics of the observed
galaxy distribution, though the color-binned 2PCF has been notoriously
difficult to correctly predict  in quantitative detail \citep[see, e.g., figures 20 \& 21 of][]{guo_etal11b}. 

There have been several recent advances in the literature on
predicting galaxy color/SFR that dramatically simplify the set of
assumptions and number of parameters required by most semi-analytic
models.  \citet{mutch_etal13} couple dark matter halo accretion
histories from the Millennium Simulation \citep{springel_etal05} with
two simple analytic functions designed to encode the physics of
baryonic growth and star formation history inside halos \citep[see also][for a closely related formulation]{cattaneo_etal11}. \citet{wang07} also use halo accretion histories in the Millennium Simulation together with a model of stellar population synthesis to paint galaxies with spectral energy distributions and star formation rates onto dark matter halos; after fitting for two quenching time parameters, their model enjoys similar success to ours at predicting two-point clustering statistics. 
\citet{peng_etal10,peng_etal12} begins with a minimal set of
phenomenological observations based on zCOSMOS data
\citep{lilly_etal07} and apply a set of continuity equations to the
evolution of the number of red and blue galaxies in order to connect
galaxy properties to the dark matter halos hosting them \citep[see also][]{lilly_etal13}. The ``separability condition" discussed in \citet{peng_etal10} is reflected by our formulation of the property $\zstarve,$ in that $\zchar,$ $\zform,$ and $\zacc$ are calculated purely independently for each halo. 
 

The studies discussed above, \citep[that is,][]{mutch_etal13,lilly_etal13,peng_etal12} assume a small set of simple analytic
formulae together with knowledge of structure formation in
$\Lambda\mathrm{CDM}$ to accurately reproduce a wide variety of
one-point statistics of the galaxy distribution. By contrast, age distribution matching uses one-point statistics as
input\footnote{Specifically the luminosity function
  $\Phi_{\mathrm{SDSS}}(L)$ and the distribution of colors $\Psdss$.} and correctly
reproduces the two-point function (Figure~\ref{fig:lum_color_wp}) as
well as higher-order statistics based on galaxy groups
(Figure~\ref{fig:color_fblue_Lbcg}). {\em The key to the
success of our theory is the identification of a special time
$\zstarve$ in each halo's assembly history that is physically associated with the slowing and ultimate quenching of star formation in the galaxy at the center of the halo.}

\subsection{Future Work}
\label{subsec:future}

The primary aim of this paper is to introduce a new theoretical
framework for studying the galaxy-halo connection, and to show that
this model is  successful even with our relatively simple
implementation. Accordingly, wherever possible
we have opted to employ simplifying approximations that are widely
used in the literature rather than tuning our exploration of halo mass
accretion history to provide better fits to the observations. For
example, as discussed in Appendix A we assume that the formation
history of dark matter halos can be encompassed by a single parameter,
$\zform,$ and we appeal to the results in \citet{wechsler02} in our
use of halo concentration as a proxy for $\zform.$ In this section we
call direct attention to some of these simplifying assumptions, and
discuss possible extensions of our model that we intend to explore in
future work. 

\subsubsection{Consideration of Time Delays}

As discussed at length in \citet{wetzel_etal11}, the robust
bimodality of the specific star formation rate (SSFR) distribution at
low-redshift constitutes compelling evidence that post-accretion
quenching of satellite star formation occurs rapidly after substantial
time delay ($\sim2-4$Gyr). This motivates a modification to the formalism we
introduce here, in which rather than $\zacc,$ the $(g-r)$ color of environment-quenched satellites 
is instead correlated with $\zquench\equiv\zacc+\Delta
t_{\mathrm{delay}}.$ We have conducted a preliminary investigation of
such models, finding that $\Delta\tdelay=2$Gyr has little effect on
either the 2PCF or the group statistics. However, the physical
processes leading to a correlation between color and $\zchar$ may also
have a delayed onset, for example due to the length of time required
for a pressure-supported shock to propagate to the virial
radius. Since the physics underlying these two delays
($\Delta\tdelay^{\mathrm{char}}$ and $\Delta\tdelay^{\mathrm{acc}}$)
is likely quite distinct, there is no reason to assume that they are
equal. We postpone a more systematic investigation of this parameter
space for the companion paper mentioned above. 

\subsubsection{Implementation of Scatter}

In  abundance matching, there is scatter between luminosity
and $\vmax.$ In this work, we have assumed a purely monotonic relation
between $(g-r)$ color and the property $\zstarve,$ but it is easy to
imagine implementations of age distribution matching in which
there is scatter in this relation as well. A simple way to accomplish
this would be an adaptation of the scatter method detailed in
Appendix A of \citet{hearin_etal12}. Briefly, one would simply
introduce Gaussian noise in the randomly drawn $(g-r)$ colors and use
the rankings of the noisy $(g-r)$ values to appropriately shuffle the
$\zstarve-$ranking of the (sub)halos. Again, we have not pursued this
complication because our present aim is to show that even the simplest
formulation of age distribution matching is remarkably
successful. However, we intend to explore such extensions in future
investigations of this technique, particularly when color-binned
galaxy-galaxy lensing measurements become available.

\subsubsection{Generalizations of the Model}

In this paper, we have only investigated our theory at low redshift.
However, if the luminosity function and color PDF are well measured at
other epochs then our model should be consistent across redshifts, and
this is a crucial next step for testing the generality of our age distribution matching formalism.  Furthermore, our formalism generalizes in the obvious way to mock galaxies with 
stellar mass and/or (S)SFR instead of luminosity. We intend to study mocks based on such properties in the future as
these are fundamental galaxy properties and are more directly linked
to color (for example, due to dust attenuation). Such tests will be enabled by existing 2PCF measurements as a function of stellar
mass split on blue and red galaxy populations at $z\sim 0$ \citep{li06,jiang_etal11a} and also $z\sim 1$ \citep{mostek12}. 
 
 Our model is also general in the sense that the first phase of the algorithm need not necessarily be  abundance matching. 
 For example, one could instead begin by using the CLF formalism to assign brightnesses to dark matter halos. Alternatively, 
 a stellar mass to halo mass map could be applied to construct a mock based on stellar masses. In either case, the color assignment phase of the algorithm 
would proceed in an identical fashion to what is described in \S~\ref{subsubsec:grassign}.

Finally, we note that in this work we have not explicitly illustrated the relationships between $\zstarve,$ halo mass, and luminosity (stellar mass), nor have we presented the relative importance of $\zacc,$ $\zform,$ and $\zchar$ as a function of halo mass and luminosity (stellar mass). Such an investigation is of direct physical interest and may yield important insights into galaxy formation and evolution. We intend to conduct a detailed study of these relationships in future work once we have thoroughly explored more complex implementations of the age distribution matching technique along with the inclusion of additional observational constraints.. 
 

\section{SUMMARY}\label{sec:conclusion}

In this section we summarize our primary results and conclusions:

\ben
\item [\textbf{1.}] We introduce {\em age distribution matching,} a new
  theoretical framework for studying the relationship between halo
  mass accretion history and stellar mass assembly.
\item [\textbf{2.}] By construction, over the entire luminosity range of our sample
  the $(g-r)$ color PDF of mock galaxies based on age distribution matching is in exact agreement with $\Psdss.$
\item [\textbf{3.}] Our model successfully predicts luminosity- and color-binned
  two-point projected galaxy clustering at low-redshift, as well as
  central and satellite galaxy color as a function of host halo
  environment ($M_{r}^{\mathrm{BCG}}$). We emphasize that these are {\em
    predictions} of the theory, and that we have not relied upon any fitting or fine-tuning of any model parameters. 
    \item [\textbf{4.}]  The success of our model implies that the assembly history of galaxies and halos are indeed correlated. 
\item [\textbf{5.}] We make publicly available  
\bit
\item Our $M_{\mathrm{r}}<-19$ mock galaxy catalog.
\item Full mass accretion histories of the main progenitors of all
  $z=0$ Bolshoi halos.  
\eit  
\een



\section{acknowledgments}

APH and DFW are particularly grateful to Andrey Kravtsov for
invaluable feedback throughout the development of this work. We
also thank Matt Becker, Benedikt Diemer, Simon Lilly, Jeff Newman, Andrew Zentner, Idit Zehavi, Jennifer Piscionere, and Nick Gnedin for useful
discussions at various stages, and Charlie Conroy and Frank van den Bosch for comments on an early draft. 
We thank Andreas Berlind for the use of his group finder and for early discussions about this type of modeling, and 
Peter Behroozi for making his
halo catalogs and merger trees publicly available, and also for
reliably prompt and clear answers to questions about these catalogs. 
We thank Ramin Skibba for sharing his mock catalog with us and for many informative discussions. 
We are grateful to an anonymous referee for valuable contributions to our first draft. 
Finally, we thank John Fahey for Transfigurations of Blind Joe Death. 

APH is also supported by the U.S. Department of Energy under contract No. DE-AC02-07CH11359.
DFW is supported by the National Science
Foundation under Award No. AST-1202698. 


\bibliography{./sham_colors}

\section*{Appendix A}

The central tenet of age distribution matching is that galaxy
color is determined by halo mass accretion history, at least in a
statistical sense.  In this appendix we discuss the details of how we
calculate $\zacc,$ $\zform,$ and $\zchar,$ the key quantities used
in the color assignment phase of our algorithm.

$\zacc:$ In accordance with the convention adopted in
\citet{rockstar_trees}, we use the snapshot after which the subhalo
always remains a subhalo. We also explored the definition of accretion
advocated in \citet{wetzel_etal13}, where subhalos are said to have
accreted the first time they have been identified as subhalos for two
consecutive time-steps, and find that this choice has little impact on
our results.

$\zform:$ We use the approximation introduced in \citet{wechsler02},
who find that mass accretion history is well-fit by an exponential,
$\Mvir(z)=M_{0}e^{-\alpha z}.$ The authors in \citet{wechsler02} defined 
the scale factor of halo formation $a_{\mathrm{form}}$ to be the epoch when 
the logarithmic slope $S$ of $\Mvir(z)$ drops below $S=2.$ While there is 
some amount of arbitrariness in this particular chosen value of $S,$ the transition that occurs 
at or near this time has very real physical consequences: the central density of the halo ceases to increase (see their Figure 18). 
This physical change naturally gives rise to a correlation between halo formation time and NFW concentration $c_{\mathrm{vir}}$  
(defined as the ratio of $\Rvir/r_{\mathrm{s}}$). 
 
 One of the central results of \citet{wechsler02} is that $a_{\mathrm{form}}$ 
is well-approximated by the following equation:
$$a_{\mathrm{form}}\equiv4.1a_{0} / c_{\mathrm{vir}},$$ 
where $\zform = 1/a_{\mathrm{form}} - 1,$ and $a_0$ is the scale factor at the time of ``observation". 
For host halos we take the time of observation to be $a_0=1$ and use the halo's concentration $c_{\mathrm{vir}}$ at this snapshot. 
Tidal disruption and mass stripping of subhalos after accretion makes
$a_0=1$ estimates of $\Rvir$ noisy and unreliable. Fortunately,
\citet{wechsler02} showed that the scaling of halo formation with
$c_{\mathrm{vir}}$ remains the same over a broad range of $a_0.$ So
for subhalos, we define $c_{\mathrm{vir}}$ to be the NFW concentration
of the subhalo at $a_0=a_{\mathrm{acc}},$ the time of accretion (as
defined above).  We note
that for $\sim10\%$ of subhalos, this definition produces an
unphysically long tail at high redshift in the $\zform$ distribution. We attribute this to
large, transient boosts to the subhalo concentration at $a_{\mathrm{acc}}$ due to a tidal
event. However, we find that our results are insensitive to this
behavior: when we collapse all objects in this tail to $\zform=3$ we
get virtually identical results.

$\zchar:$ We use the first time a (sub)halo attains $10^{12}\hMsun.$
As noted in the main body of the paper, we explored a range of values
of $\mchar$ and found $10^{12}\hMsun$ is the most successful.


\end{document}